\documentclass[%
 reprint,
 amsmath,amssymb,
 pra,
]{revtex4-1}

\pdfoutput=1

\usepackage{graphicx}
\usepackage{dcolumn}
\usepackage{bm}
\usepackage{amsmath}
\usepackage{esvect}
\usepackage{esdiff}
\usepackage{placeins}
\usepackage{afterpage}
\usepackage{xcolor}
\usepackage{nomencl}
\usepackage[acronym]{glossaries}
\usepackage{siunitx}
\usepackage{bm}
\usepackage{esvect}
\usepackage{color}

\DeclareMathOperator{\chitwo}{ \chi^{(2)} }

\newcommand{\f}{f}
\newcommand{\s}{s}
\newcommand{\su}{ {s1} }
\newcommand{\st}{ {s2} }
\newcommand{\om}{ \omega}
\newcommand{\dbetas}{ \Bar{ \beta}_{s}^{(1)} } 
\newcommand{\dbetaf}{ \Bar{ \beta}_{f}^{(1)} } 
\newcommand{\mucw}{ \mu_c } 
\newcommand{\musin}{ \Bar{ \beta}_{sin} }
\newcommand{\eps}{\varepsilon}

\newcommand{\eq}[1]{
\begin{equation}\begin{aligned}
#1
\end{aligned}\end{equation}}

\begin{document}

\preprint{APS/123-QED}

\title{  Solitons near avoided mode crossing in $\chitwo$ nanowaveguides }

\author{William R. Rowe}
\author{Andrey V. Gorbach}%
 \email{A.Gorbach@bath.ac.uk}
\author{Dmitry V. Skryabin}
\affiliation{Department of Physics, University of Bath, Bath BA2 7AY, England, UK\\
Centre for Photonics and Photonic Materials, University of Bath, Bath BA2 7AY, England, UK
}%

\date{\today}

\begin{abstract}
We present a model for $\chitwo$ waveguides accounting for three modes, two of which make an avoided crossing at the second harmonic wavelength. We introduce two linearly coupled pure modes and adjust the coupling to replicate the waveguide dispersion near the avoided crossing. Analysis of the nonlinear system reveals continuous wave (CW) solutions across much of the parameter-space and prevalence of its modulational instability. We also predict the existence of the avoided-crossing solitons, and study peculiarities of their  dynamics and spectral properties, which include formation of a pedestal in the pulse tails and associated pronounced spectral peaks. Mapping these solitons onto the linear dispersion diagrams, we make connections between their existence and CW existence and stability. We also simulate the two-color soliton generation from a single frequency pump pulse  to back up its formation and stability properties. 
\end{abstract}

\pacs{Valid PACS appear here}
\maketitle


\section{Introduction}
Interest in temporal quadratic ($\chitwo$) solitons has seen recent resurgence \cite{Kowligy2018Mid-infraredWaveguides, Guo2015SupercontinuumMatching, Rowe2019TemporalNanowaveguides, Rowe2020RamanNonlinearities} thanks to the development of nano-waveguides over the last decade \cite{Poberaj2012LithiumDevices, Boes2018StatusCircuits, Huang2015Sub-micronLithography, Wilson2020IntegratedPhotonics}. In particular, Lithium Niobate (LiNbO$_3$, LN) nano-waveguides provide the strong $\chitwo$ response and broadband transparency required for a range of applications including supercontinuum generation (SCG) and pulse compression \cite{Zeng2008Two-colorNiobate, Bache2010Type-ILasers, Zeng2012SolitonNonlinearities, Guo2015SupercontinuumMatching, Yu2020UltravioletWaveguides,Wang2018Ultrahigh-efficiencyWaveguides,Lu2019EfficientNanophotonics, Zhao2020Shallow-etchedGeneration}. These nano-structures, fabricated from single crystals, allow for compact, low loss waveguides with $\chitwo$ nonlinearity that weren't previously feasible \cite{Wang2018Ultrahigh-efficiencyWaveguides,  Zhao2020Shallow-etchedGeneration, Lu2019EfficientNanophotonics}.

Solitons along with their interaction with dispersive waves have been found to play a significant role in various frequency conversion processes \cite{Skryabin2010TheoryWaves, Rowe2019TemporalNanowaveguides, Rowe2020RamanNonlinearities}. For efficient soliton generation phase and group velocity matching between the fundamental and second harmonic wavelengths are required \cite{Rowe2019TemporalNanowaveguides}. Quasi-phase matching (QPM) can be achieved with periodic polling, a technique which is well understood in LN \cite{Poberaj2012LithiumDevices, Wang2018Ultrahigh-efficiencyWaveguides, Zhao2020Shallow-etchedGeneration}. Group velocity matching (GVM) in LN, however, is more difficult to achieve. It is well known that bulk LN has the zero of its group velocity dispersion (GVD) around $\lambda=1.9\si{ \micro\metre }$ \cite{Bache2010Type-ILasers, Zelmon1997InfraredNiobate}, making GVM between wavelengths in the near-infrared and visible range impossible. This is where the strong geometric dispersion of LN nano-waveguides can be used to  shift the zero-GVD point to shorter wavelengths making GVM possible between the desired frequencies \cite{Zhu2021IntegratedNiobate}.

Another powerful technique of arranging GVM has been recently proposed for LN nano-structures \cite{Cai2018HighlyNano-waveguides}, 
which relies on engineering of an avoided crossing between different guided modes. Strong modification of dispersion induced by avoided crossings is known to impact existence and properties of solitons. Solitons spectrally centred near an
avoided crossing and also the impact of the avoided crossing on the soliton detuned far away from it have been previously studied in photonic crystal  fibers 
 and  microresonators, see, e.g., \cite{Skryabin2004CoupledFibers, Tani2018EffectFibers} and
\cite{Yulin2005DissipativeFilms, Liu2014InvestigationGeneration,Herr2014ModeMicroresonators,Wang2020DiracMicroresonators}, respectively.

Typically, avoided mode crossings occur as the result of interactions between guided modes supported by different sub-components of a complex structure \cite{Cai2018HighlyNano-waveguides, Skryabin2004CoupledFibers}, or interactions between forward- and backward-propagating modes induced by Bragg gratings \cite{Skryabin2004CoupledFibers,Yulin2005DissipativeFilms}. In microresonators, avoided crossings between different mode families can also be induced by geometrical imperfections \cite{Liu2014InvestigationGeneration}. In our study we find avoided-crossings between quasi-TE and quasi-TM guided optical modes of LN nano-waveguides. These avoided-crossings appear to be induced by high anisotropy: the combined effect of the intrinsic material anisotropy of LN and structural properties.  

\begin{figure*}
    \centering
    \includegraphics{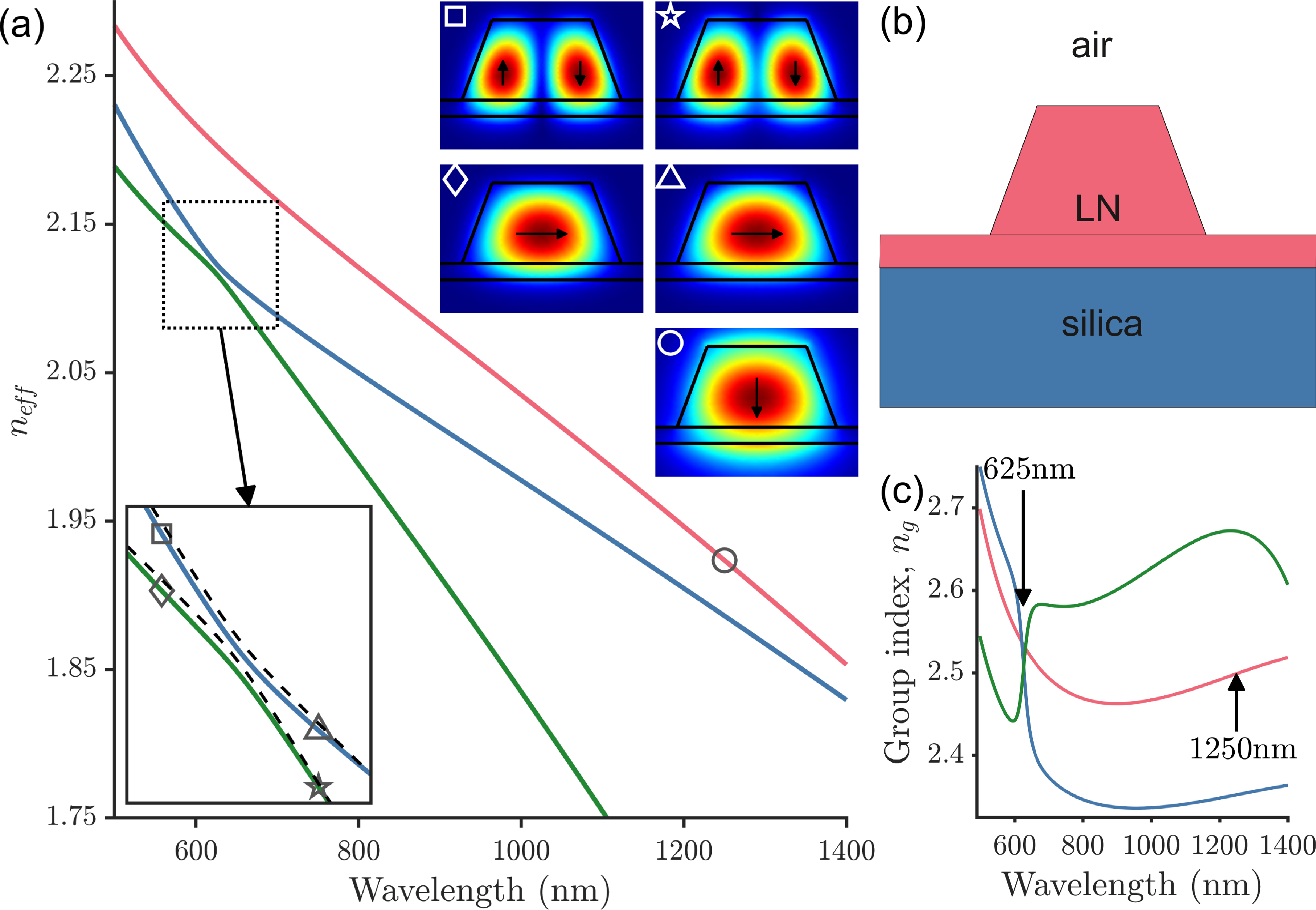}
    \caption{ Modelled (a) effective refractive indices ($n_{e\!f\!f}$) and (c) group indices $n_g$ of three guided modes of (b) LN nano-waveguide structure (top width 600nm, ridge height 500nm and slab thickness 100nm, X-cut LN is considered with the extraordinary axis oriented vertically on the diagram). Fundamental TM mode shown in red, two hybridised modes plotted in blue and green (transverse profiles swap due to hybridisation). Black dotted lines plot $n_{e\!f\!f}$ produced by our model Eq. \eqref{c:mode_crossing:e:betapm}. Symbols mark the positions of the corresponding traverse mode profile insets (arrows show dominant polarisation). }
    \label{c:mode_crossing:f:neff}
\end{figure*}

Fig. \ref{c:mode_crossing:f:neff}(a) shows one example of an avoided-crossing between two modes of the LN waveguide structure shown in Fig.~\ref{c:mode_crossing:f:neff}(b). 
Due to the anisotropy, at short wavelengths higher-order quasi-TM modes appear to have higher effective indexes than the fundamental quasi-TE mode. At long wavelengths, however, as the modes become less localized and the anisotropy is effectively suppressed, both fundamental polarizations appear to have larger indexes than any other higher-order modes. This generally results in avoided-crossings between different pairs of quasi-TM and quasi-TE modes. 

The strong induced dispersion at the anti-crossings causes an abrupt change in the group index of each mode, allowing GVM between different guided modes across the optical octave (1250nm - 625nm in our example), as shown in Fig.~\ref{c:mode_crossing:f:neff} (c). 
Assuming in addition a suitable QPM is arranged, we investigate solitons emerging from nonlinear interactions between the two avoided-crossing modes in the visible (second harmonic frequency, SH) and the corresponding phase- and group velocity matched mode in the infrared (fundamental frequency, FF).

Previous work has been done in mathematically similar systems in which Bragg gratings couple forward and backward traveling waves, producing simultaneously gaps in the linear spectrum of FF and SH fields \cite{Conti1998TrappingSolitons, Arraf2001BrightNonlinearity, Leitner2005StabilityNonlinearity}. Slow $\chitwo$ solitons (with group velocities close to zero physical velocity) are known to exist in such systems. The system we investigate here is similar in that the avoided-crossing provides a gap in the linear spectrum of the SH. Notably, the avoided crossing is produced by two co-propagating modes in our case, and there is only one mode in the FF field. Any solitons will therefore be fast (close to the speed of light). 

Another important feature of our system is the change in the transverse profiles of the modes across the avoided-crossing, associated with polarization rotation. The insets in Fig.~\ref{c:mode_crossing:f:neff}(a) show how the profiles of the modes swap from quasi-TM to quasi-TE (or vice versa) on either side of the avoided crossing. Effective nonlinear interaction between FF and SH modes depends on the spatial overlap between the relevant modes \cite{Cai2018HighlyNano-waveguides, Rowe2019TemporalNanowaveguides}.
We therefore see a dramatic dispersion of the effective nonlinearity between FF and SH modes, which plays a significant role in the properties of CW solutions and solitons.

\section{Model}

We proceed with presenting a model of interacting three modes: a single FF mode centered around a reference frequency $\omega_f$, and two SH avoided crossing modes centered around the frequency $\omega_s=2\omega_f$. For the example presented in Fig.~\ref{c:mode_crossing:f:neff} we select $\lambda_f=2\pi c/\omega_f=1250$nm and $\lambda_s=2\pi c/\omega_s=625$nm, here $c$ is the speed of light.

\subsection{Linear dispersion}

To model dispersion of the FF mode we fit its propagation constant using the Taylor expansion
\eq{\label{c:mode_crossing:e:betaf}
\beta_\f(\delta) = \beta_\f^{(0)} + \beta_\f^{(1)} \delta + \frac{1}{2}\beta_\f^{(2)} \delta^2,
}
where $\delta = \om - \om_\f$ is the frequency detuning centred from the reference frequency $\om_\f$. For our example in Fig.~\ref{c:mode_crossing:f:neff} we find $\beta_\f^{(0)} = \SI{9700}{mm^{-1}} $, $\beta_\f^{(1)} = \SI{8.3}{ps/mm} $ and $\beta_\f^{(2)} = -3.6 \times 10 ^{-4}\si{ps^2/mm} $.

For the SH avoided crossing modes, instead of fitting directly their dispersion, we adapt the well-known linear coupler model. The advantage of this approach will become apparent in the following sub-section, where we introduce nonlinear couplings between the modes. 

Unlike in many typical avoided crossing setups such as evanescently coupled waveguides or Bragg gratings (coupled forward- and backward-propagating modes), there is no natural basis of uncoupled modes in our system. 
We introduce two artificial "pure" modes SH1 and SH2
by fitting the dispersion of the true guided modes away from the avoided crossing region, which provides propagation constants crossing at around 625nm: \eq{\label{c:mode_crossing:e:betasm}
\beta_{s m}(\delta)=\beta_{s}^{(0)} + 2\delta\left(\beta_{s}^{(1)}+(-1)^{m} \dbetas \right), ~m=1,2.
}
Here $2\delta = \omega - \omega_s$ with $\omega_s = 2\omega_\f$ are the SH frequency detunings about the SH frequency. For our example $\beta^{(0)}_s=2.2 \times 10^{4} \si{mm^{-1}}$, $\beta^{(1)}_s=\SI{8.4}{ps/mm }$ and   $\dbetas=\SI{0.36}{ps/mm}$ and accounts for the difference in group velocities of the SH modes.

The avoided-crossing effect is then introduced via a linear coupling, $C$, between SH1 and SH2:
\eq{\label{eq_coupler}
-i\partial_z \begin{bmatrix}a_{s1} \\ a_{s2}\end{bmatrix}  = \begin{bmatrix} \beta_{s1}(\omega)&  C \\  C&  \beta_{s2}(\omega) \end{bmatrix} \begin{bmatrix}a_{s1} \\ a_{s2}\end{bmatrix} .
}
Diagonalising this system gives propagation constants for the avoided crossing modes SH+ and SH- as
\eq{\label{c:mode_crossing:e:betapm}
\beta_{s\pm} = \frac{\beta_{s1} +\beta_{s2}}{2} \pm \sqrt{ \frac{(\beta_{s1} -\beta_{s2})^2}{4} +C^2 }.
}
The coupling coefficient $C$ can be obtained by fitting the above propagation constants $\beta_{s+}$ and $\beta_{s-}$ to the actual dispersion of guided modes in the avoided crossing region. The inset of Fig.~\ref{ c:mode_crossing:f:sol_sims }(a) shows the comparison between the actual dispersion (solid lines) and dispersion provided by the model in Eq.~\eqref{c:mode_crossing:e:betapm} with $C=34\si{mm^{-1} }$.
Close to the avoided crossing, GVD and higher-orders of dispersion in the SH+ and SH- modes is dominated by the mode anti-crossing effect, which allows us to limit terms in Eq.~\eqref{c:mode_crossing:e:betasm} to first order.

\subsection{Nonlinear envelope equations}

We complete our model by adding the $\chitwo$ nonlinear coupling between FF and both SH modes. This gives the envelopes of the fundamental, $F_\f$, and two second harmonic, $F_\su$, $F_\st$ fields satisfy
\eq{\label{c:mode_crossing:e:model}
i \partial_z F_\f + (i \dbetaf \partial_\tau - \frac{1}{2} \beta_\f^{(2)} \partial_\tau^2 ) F_\f + \gamma F_\f^* ( F_\su + \alpha F_\st ) = 0, \\
i \partial_z F_\su + (\kappa - i \dbetas \partial_\tau ) F_\su + C F_\st + \frac{\gamma}{2} F_\f^2  = 0, \\
i \partial_z F_\st + (\kappa + i \dbetas \partial_\tau ) F_\st + C F_\su + \frac{\gamma\alpha}{2} F_\f^2  = 0, \\
}
Here $z$ is the coordinate along the waveguide, $\tau=t-z\beta_\s^{(1)}$ is the retarded time with $t$ being the physical time and $\beta_\s^{(1)}$ is the average inverse group velocity of the SH modes, cf. Eq.~\eqref{c:mode_crossing:e:betasm}, $\dbetaf=\beta_\f^{(1)}-\beta_\s^{(1)}$ is the FF inverse group velocity in this moving frame ($\dbetaf = \SI{0.086}{ps/mm}$ for our example). The amplitudes, ($F_\f$, $F_\su$ and $F_\st$) are measured in the units of $\sqrt{W}$, and the respective electric fields are $F_\f e^{i(z\beta_\f^{(0)}-\om_\f t)}+c.c.$, and
$F_{\s m} e^{i2(z\beta_\f^{(0)}-\om_\f t) }+c.c.$.

The QPM period, $2\pi/G$, is assumed to set the phase-mismatch between the FF and SH modes. This makes the phase mismatch parameter $\kappa=\beta_\s^{(0)}-2\beta_\f^{(0)}+G$ which is identical for both SH pure modes as we set $\om_\s$ as the frequency at the centre of the mode crossing. In the examples below we will closely examine the case of exact phase matching, $\kappa=0$. We retain $\kappa$ in the model equations to keep our analysis general.

The nonlinear interaction strength between the FF mode and each of the pure modes SH1 and SH2 is given by the coefficients $\gamma$ and $\gamma\alpha$, respectively. Without loss of generality we assume $0\le\alpha\le 1$. Coefficients $\gamma$ and $\alpha$ can be estimated from the modal overlaps using the actual SH guided modes of the structure away from the avoided crossing region. The dispersion of nonlinearity associated with the pronounced reshaping of the modes (polarization rotations) in the avoided crossing region is fully incorporated in our model by virtue of the structure of the eigen-vectors of the linear coupler system in Eq.~\eqref{eq_coupler}. Thus, in the limit of $\alpha=0$ the effective interaction strength between FF and each SH+ and SH- mode varies between zero and full strength across the avoided crossing region. In the opposite limit of $\alpha=1$ the interaction remains constant.
For the example geometry shown in Fig.~\ref{c:mode_crossing:f:neff}(b) we obtain $\gamma=\SI{400}{m^{-1}W^{-1/2} } $  and $\alpha=0.14$.

\section{Continuous wave solutions}

To analyse continuous wave (CW) solutions of Eq. \eqref{c:mode_crossing:e:model} we make the substitutions
\eq{\label{c:mode_crossing:e:CW_ansatz}
F_\f = \Tilde{A}_\f e^{i\mucw z - i \delta \tau},\\
F_{\s m } = \Tilde{A}_{\s m } e^{i2\mucw z - i 2\delta \tau},\\
}
where $m=1,2$, $\mucw$ is the propagation constant for CW solutions and $\Tilde{A}_\f$ and $\Tilde{A}_{\s m }$ are the CW solution amplitudes for the FF and SH modes respectively. Taking this anzats in the low amplitude limit and linearizing Eq.\eqref{c:mode_crossing:e:model} we obtain propagation constants for FF and SH modes respectively as,
\eq{
\mucw \rightarrow \Bar{\beta}_\f = \dbetaf \delta + \frac{1}{2} \beta_\f^{(2)} \delta^2,\\
2\mucw \rightarrow  \Bar{\beta}_{\s \pm} = \frac{\kappa}{2} \pm \sqrt{ \frac{C^2}{4}+ (\delta \dbetas)^2 },
}
which are the analogues of the linear dispersions in Eqs.~\eqref{c:mode_crossing:e:betaf} and \eqref{c:mode_crossing:e:betapm}
in the rotating and moving reference frame of our model in Eq.~\eqref{c:mode_crossing:e:model}. 

\begin{figure*}
    \centering
    \includegraphics{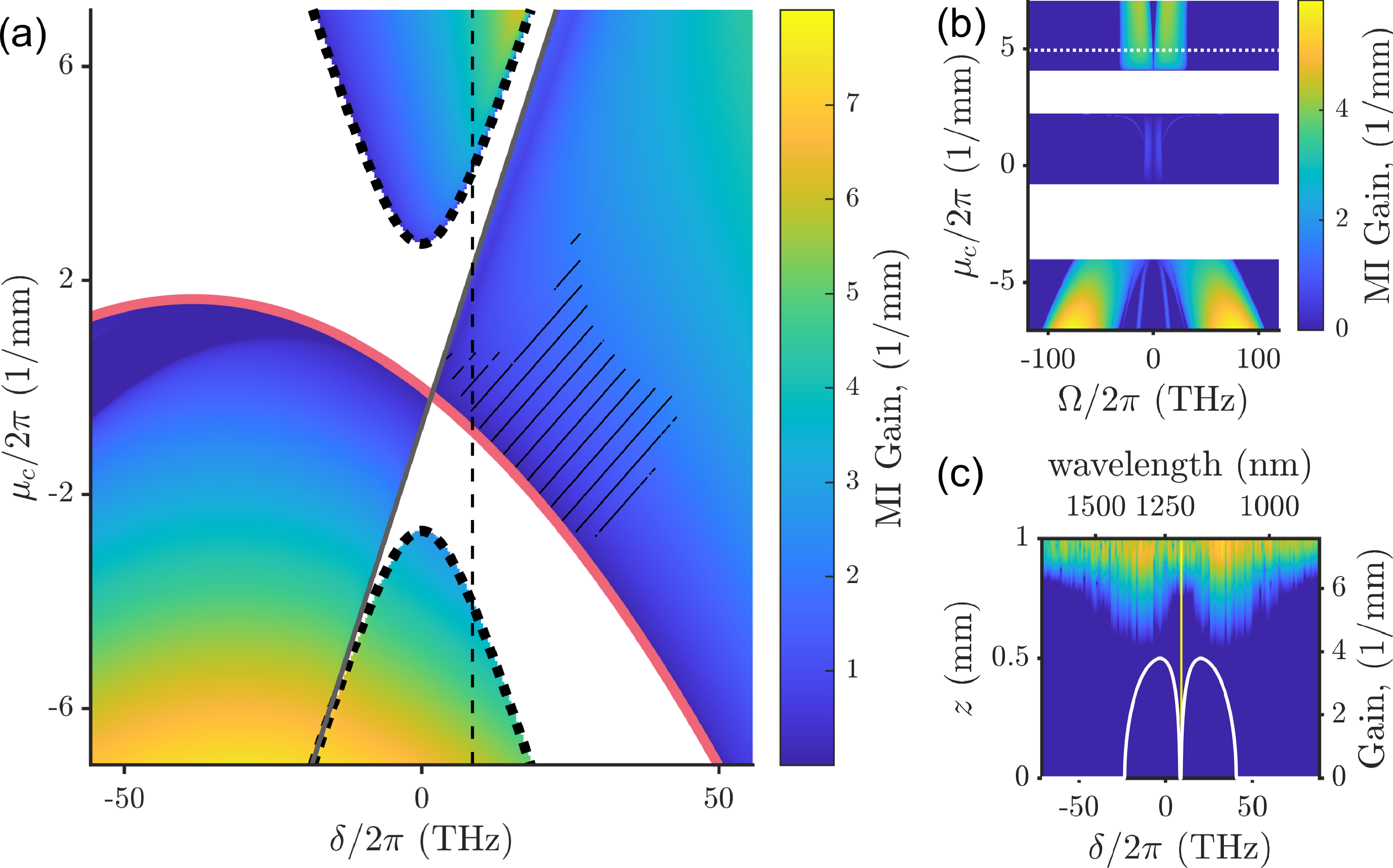}
    \caption{ (a) Linear dispersion of FF and SH modes, $\mu_c = \beta_\f$ and $\mu_c = \beta_\pm$ (thick red (solid) and black (dotted) lines respectively) and line of $\mu_c=\musin$ shown in grey. CW solutions exist in the coloured regions and the colour scale marks the maximum MI gain. Soliton solutions have been found with the FF frequencies of their peaks in time in the region hatched by the thin black lines. (b) Full MI gain spectrum for $\delta = \SI{8.5}{THz}$ as marked by the vertical (dashed) line in (a). (c) Spectrum of FF component from simulated propagation of  CW solution of $\mucw = \SI{4.9} mm^{-1}$, $\delta=\SI{8.5}{THz}$. Predicted MI gain plotted in white corresponding to white (dotted) horizontal line in (b). All data for case of exact phase matching ($\kappa=0$). }
    \label{c:mode_crossing:f:cw}
\end{figure*}

We then find nonlinear CW solutions of the form:

\begin{eqnarray}
\label{eq_Af_cw}
\gamma^2|\Tilde{A}_\f|^2 &=& \frac{(\mucw - \Bar{\beta}_\f)(2\mucw - \Bar{\beta}_{\s+})(2\mucw - \Bar{\beta}_{\s-})}{ (1+\alpha^2)(\mucw - \musin) } ,\\
\Tilde{A}_\su &=& \frac{\gamma \Tilde{A}_\f^2 }{2} \frac{ (2\mucw - \kappa - 2\dbetas\delta ) + \alpha C }{  (2\mucw - \Bar{\beta}_{s+}) (2\mucw - \Bar{\beta}_{s-})  }, \\
\Tilde{A}_\st &=& \frac{\gamma \Tilde{A}_\f^2 }{2} \frac{ \alpha(2\mucw - \kappa + 2\dbetas\delta ) + C }{ (2\mucw - \Bar{\beta}_{s+}) (2\mucw - \Bar{\beta}_{s-})  }.
\end{eqnarray}

where we have defined the propagation constant at which $\Tilde{A}_\f$ becomes singular as
\eq{
\musin = \frac{\kappa}{2} + \frac{  \dbetas\delta  (1-\alpha^2 )  }{1+\alpha^2} - \frac{\alpha C }{1+\alpha^2}.
}
We point out that the requirement of the r.h.s. of Eq.~\eqref{eq_Af_cw} to be non-negative defines the domains of existence of CW solutions in the 
$\mucw$-$\delta$ plane. The lines $\mucw = \Bar{\beta}_\f$, $2\mucw = \Bar{\beta}_{\s\pm}$ and $\mucw = \musin$ mark the boundaries of the regions of CW solution existence, as illustrated in  Fig.~\ref{c:mode_crossing:f:cw}(a). 

\subsection{Modulation instability}

To analyse modulation instability (MI) of the CW solutions we update our ansatz in Eq.~\eqref{c:mode_crossing:e:CW_ansatz} to
\eq{
F_\f = (\Tilde{A}_\f  + \eps_{\f\s}e^{i\Omega\tau}  + \eps_{ \f a }e^{-i\Omega\tau}  )e^{i\mucw z - i \delta \tau},\\
F_{\s m } = (\Tilde{A}_{\s m } + \eps_{\s m \s}e^{i2\Omega\tau}  + \eps_{ \s m a }e^{-i2\Omega\tau}   )e^{i2\mucw z - i 2\delta \tau},\\
}
allowing for small perturbations $\eps$ detuned by frequency $\Omega$. The final subscripts $\s$ and $a$ denote stokes and anti-stokes detuned waves respectively. Keeping terms linear in $\eps$ we form the matrix equation $\partial_z \Vec{\eps} = \bm{Q}\Vec{\eps}$ where
\begin{widetext}
\eq{
\Vec{\eps} = [\eps_{\f\s}, \eps_{\f a}^*,  \eps_{\s 1 \s}, \eps_{ \s 1 a }^*,  \eps_{\s 2 \s}, \eps_{ \s 2 a }^*  ]^\text{T},\\
\bm{Q} = i
\begin{bmatrix}
B_\f^{(-)} & \gamma( \Tilde{A}_{\su} + \alpha \Tilde{A}_\st ) & \gamma \Tilde{A}_\f^* & 0 & \gamma \alpha \Tilde{A}_\f^* & 0 \\
-\gamma( \Tilde{A}_{\su}^* + \alpha \Tilde{A}_\st^* ) & - B_\f^{(+)} & 0 & -\gamma \Tilde{A}_\f &  0 & -\gamma \alpha \Tilde{A}_\f \\
 \gamma  \Tilde{A}_\f & 0 & B_\s^{(-)} & 0 & C & 0 \\
0 & -\gamma  \Tilde{A}_\f^*  & 0 & -B_\s^{(+)} & 0 & -C \\
 \gamma\alpha  \Tilde{A}_\f & 0 & C & 0 & B_\s^{(-)} & 0  \\
0 & -\gamma \alpha \Tilde{A}_\f^*  & 0 & -C &  0 & -B_\s^{(+)} 
\end{bmatrix},\\
B_\f^{(\pm)} =  - \mucw + \dbetaf (\delta \pm \Omega) + \frac{1}{2} \beta_\f^{(2)}(\delta \pm \Omega)^2 , \\
B_\s^{(\pm)} = \kappa - 2\mucw - 2 \dbetas (\delta \pm \Omega).
}
\end{widetext}
The real parts of the eigenvalues of $\bm{Q}$ give the MI gain in the system. Numerically computing the MI gain around the SH mode anti-crossing we find it is non-zero across most of the region. The maximum MI gain is plotted in Fig.~\ref{c:mode_crossing:f:cw}(a) across the $\mucw$-$\delta$ plane. The MI gain structure in $\Omega$ was generally found to have between two and six peaks. Examples of this are shown in Figs.~\ref{c:mode_crossing:f:cw}(b) and (c). We verified our MI predictions by comparison with numeric simulations of CW solution propagation with the addition of low level white noise to seed the MI. One representative example of this is given in Fig.~\ref{c:mode_crossing:f:cw}(c), where we can see new frequencies emerging in the simulation in the regions which coincide with those predicted by our MI gain analysis.

\begin{figure*}
    \centering
    \includegraphics{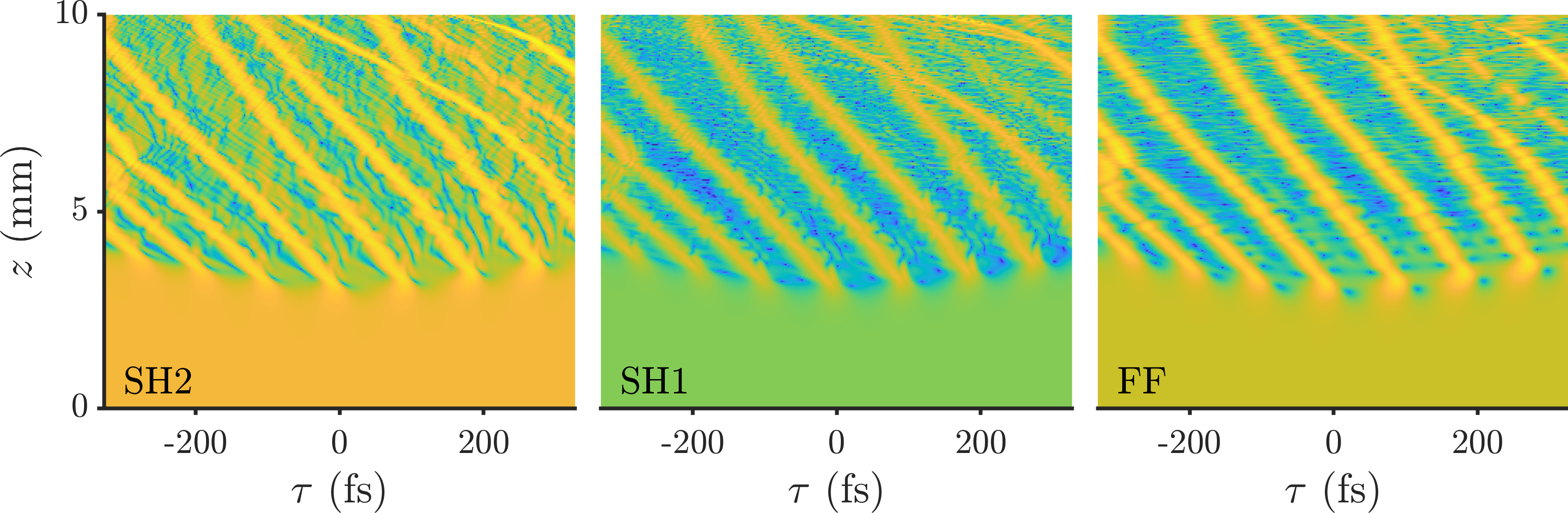}
    \caption{ Propagation simulation of CW solution ($\mucw = \SI{0.35}{mm^{-1}}$, $\delta=\SI{8.5}{THz}$) undergoing MI and subsequent formation of solitons. Panels show $\ln{|F_\st|}$, $\ln |F_\su|$ and $\ln | F_\f|$ from left to right in the time domain. Data for case of exact phase matching ($\kappa=0$).  }
    \label{c:mode_crossing:f:CW_prop}
\end{figure*}

Some simulations of MI were found to produce solitonic pulses after sufficient propagation distance, see an example in Fig.~\ref{c:mode_crossing:f:CW_prop}. These were easily identified as they did not undergo dispersion and all three components propagated together, not at their respective group velocities. This group velocity locking is a clear signature of sustained nonlinear interactions between the three components.

\section{Solitons}

To analyse the existence of solitons in the system we make a new substitution into Eqs.~\eqref{c:mode_crossing:e:model} to allow for time dependent solutions
\eq{
F_\f = A_\f(\eta) e^{i\mu z},\\
F_{sm} = A_{sm} (\eta) e^{2i\mu z} ,
}
where we introduce the reference frame of the soliton with an inverse velocity $\nu$ by defining $\eta=\tau - \nu \xi$. This gives our model for soliton envelope functions as,
\begin{widetext}
\eq{\label{c:mode_crossing:e:soliton_model}
(- \mu + i (\dbetaf - \nu) \partial_\eta - \frac{1}{2} \beta_\f^{(2)} \partial_\eta^2 ) A_\f + \gamma A_\f^* ( A_\su + \alpha A_\st ) = 0, \\
(\kappa-2\mu - i (\dbetas+\nu) \partial_\tau ) A_\su + C A_\st + \frac{\gamma}{2} A_\f^2  = 0, \\
(\kappa-2\mu + i (\dbetas-\nu) \partial_\tau ) A_\st + C A_\su + \frac{\gamma\alpha}{2} A_\f^2  = 0. \\
}
\end{widetext}
Thus the soliton parameters $\mu$ and $\nu$ uniquely define the soliton solutions. The frequency shift of the solitons is a function of $\mu$ and $\nu$ and is independent for the three field components, it is therefore incorporated into the definitions of $A_\f$ and $A_{\s m}$. We note that we recover our model for CW solutions by setting $A_f(\eta) = \Tilde{A}_f e^{-i\delta\eta}$ and $A_{sm}(\eta) = \Tilde{A}_{sm} e^{-2i\delta\eta}$ which shows us that soliton and CW propagation constants are related by 
\eq{\label{c:mode_crossing:e:mucw_def}
\mucw = \mu + \nu \delta.
}

\subsection{Localisation Analysis}

\begin{figure*}
    \centering
    \includegraphics{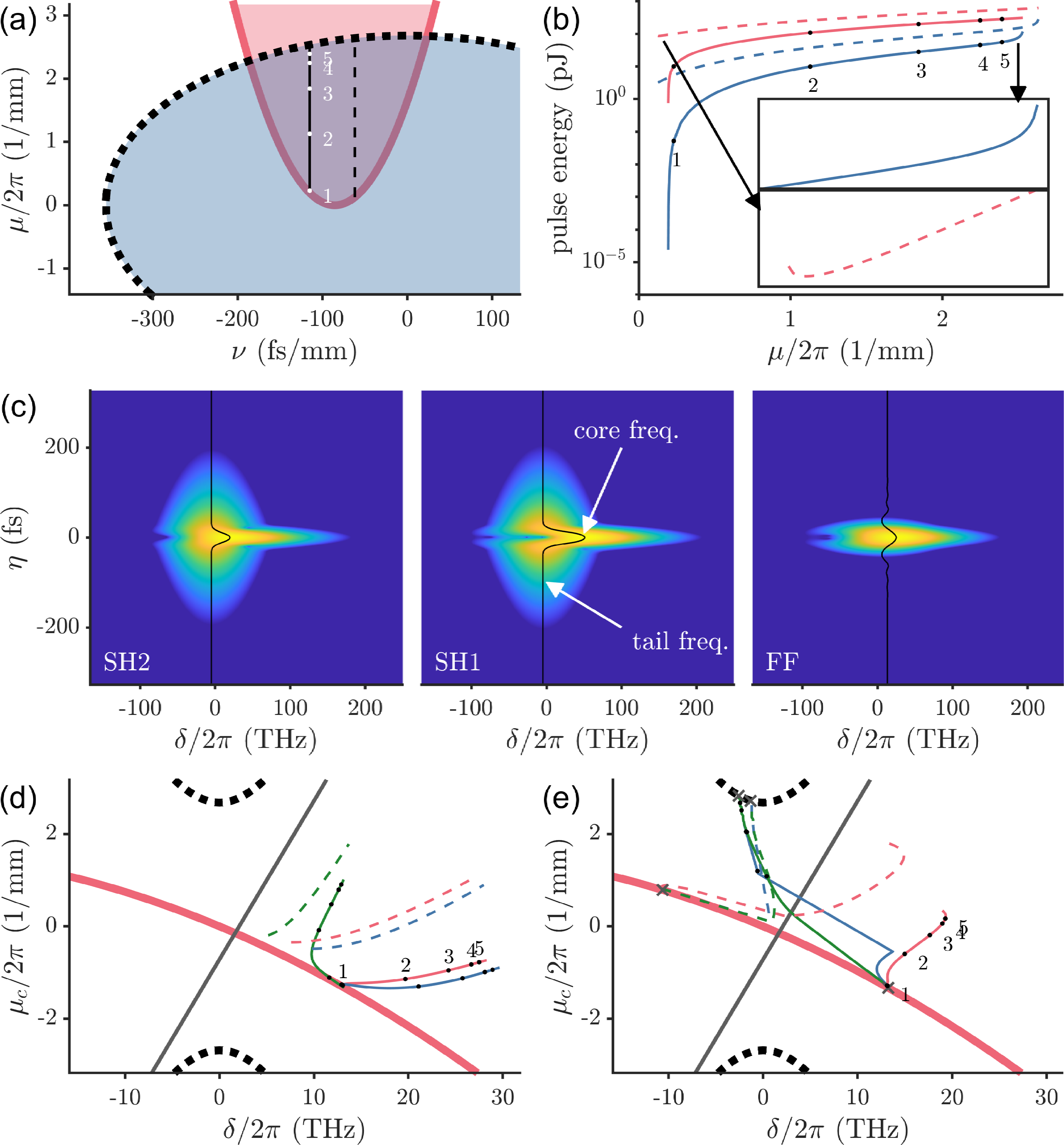}
    \caption{ Soliton analysis for the case of exact phase matching ($\kappa=0$). 
    (a) Soliton existence criteria plotted on $\mu$-$\nu$ plane shown as shaded red and blue areas for the FF and SH localisation respectively. Vertical solid and dashed lines plot constant $\nu=-\SI{115}{fs/mm}$ and $\nu=-\SI{62}{fs/mm}$  trajectories respectively.
    (b) Energy in each soliton component as a function of $\mu$. Solid and dashed lines correspond to the same trajectories as in panel (a). Insets give zoomed view of low and high $\mu$ ends of FF and SH trajectories respectively. Trend for SH2 follows SH1 closely and is omitted for clarity. 
    (c) XFROG plot of soliton 5 as marked in the other panels in this figure calculated numerically using Newton Raphson method. Black curves mark the peak frequency in each component as a function of time. Gaussian reference pulse used for XFROG had a FWHM of 24fs.
    (d) and (e) show each soliton component plotted on the $\mucw$-$\delta$ plane, thick curves are as in Fig.~\ref{c:mode_crossing:f:cw}(a). Frequencies, $\delta$, in (d) and (e) are from the temporal peak and tail respectively of the solitons. Thin lines correspond to same trajectories as in panel (a). Red, blue and green lines correspond to FF, SH1 and SH2 components respectively. Dots labeled with numbers correspond positions of examples solitons shown in panel (c) and Fig.~\ref{c:mode_crossing:f:soliton_profiles}. Grey crosses in (e) mark soliton tail positions according to Eq.~\eqref{c:mode_crossing:e:delta_tails}. }
    \label{c:mode_crossing:f:soliton_analysis}
\end{figure*}

From here we analyse the system to identify criteria for localised solution existence. To do this we require that far from their centre, the soliton envelope functions decay exponentially:
\eq{
A_\f(\eta\to\pm\infty)  =  a_\f e^{-\lambda_\f |\eta|},\\
A_{\s m}(\eta\to\pm\infty) =  a_{\s m} e^{-\lambda_\s |\eta|}.
}
Linearizing the system in Eq.~\eqref{c:mode_crossing:e:soliton_model} for small amplitude soliton tails, we obtain $\lambda_f$ and $\lambda_sm$ as functions of the soliton parameters $\mu$ and $\nu$. For $\lambda_\f$ and $\lambda_\s$ to provide exponential decay we require that they have non-zero real parts, from which we derive:
\eq{\label{c:mode_crossing:e:soliton_criteria}
\mu < -\frac{(\dbetaf - \nu)^2 }{2 \beta_\f^{(2)} }  ,\\
 2\mu > \kappa - C\sqrt{1 - \bigg( \frac{\nu}{\dbetas} \bigg)^2 } ,\\
 2\mu < \kappa + C\sqrt{1 - \bigg( \frac{\nu}{\dbetas} \bigg)^2 },
}
as the conditions on localised soliton existence. These conditions are visualised on the $\mu$-$\nu$ plane in Fig.~\ref{c:mode_crossing:f:soliton_analysis}(a) for our example geometry with $\kappa=0$. Localization in all three components is possible where the different shaded regions overlap.

Extending this analysis to the imaginary parts of $\lambda_f$ and $\lambda_s$ at the soliton existence boundaries we obtain expressions for frequency detuning of the soliton tails,
\eq{\label{c:mode_crossing:e:delta_tails}
\delta_{f,tail} = \frac{\nu - \dbetaf}{\beta_f^{(2)} },\\
\delta_{s,tail} = \frac{ C }{ 2\dbetas } \bigg( \bigg( \frac{\dbetas}{\nu} \bigg)^2 - 1 \bigg)^{-1/2} .
}
This analysis is equivalent to finding the frequency at which the soliton inverse velocity and linear group velocity match as the soliton approaches the linear dispersion.

\subsection{ Numerical solutions }

Using the Newton-Raphson method we were able to find soliton solutions numerically across the entire region of their existence, as predicted by our tail analysis in Eq.~\eqref{c:mode_crossing:e:soliton_criteria} 
To analyse the structure of soliton solutions, we find it most instructive to construct cross-correlation frequency resolved optical gating (XFROG) spectrograms of the solitons. This is a commonly used technique which allows us to view both temporal and spectral structure in the soliton simultaneously. Here we produce XFROGs using
\eq{
I(t, \omega) = \text{ln} \left| \int^{+\infty}_{-\infty} \text{d}\tau' A_{ref}(\tau' - \tau) F(\tau') e^{-i \omega \tau'}\right|,
}
where $A_{ref}$ is a Gaussian reference pulse envelope, and $F$ is one of the FF, SH1 or SH2 field envelope. Using the XFROGs we observed that soliton structure varies significantly across the existence domain and many of the solitons have a pronounced frequency chirp, such that frequencies in their tails ($|\tau|\gg0$) different from their main core ($t=0$). These frequency differences are also seen to vary across the three components of the soliton. An XFROG spectrogram showing an example of such a soliton is shown in Fig.~\ref{c:mode_crossing:f:soliton_analysis}(c). Calculating soliton profiles across their whole existence domain we use Eq.~\eqref{c:mode_crossing:e:mucw_def} with the FF core frequency of the soliton to map the soliton existence domain onto the linear dispersion plot in Fig.~\ref{c:mode_crossing:f:cw}(a). We observe that the solitons exist in a region of CW solution existence, which is consistent with our earlier observations for two-component $\chitwo$ solitons \cite{Rowe2019TemporalNanowaveguides}.

\begin{figure*}
    \centering
    \includegraphics{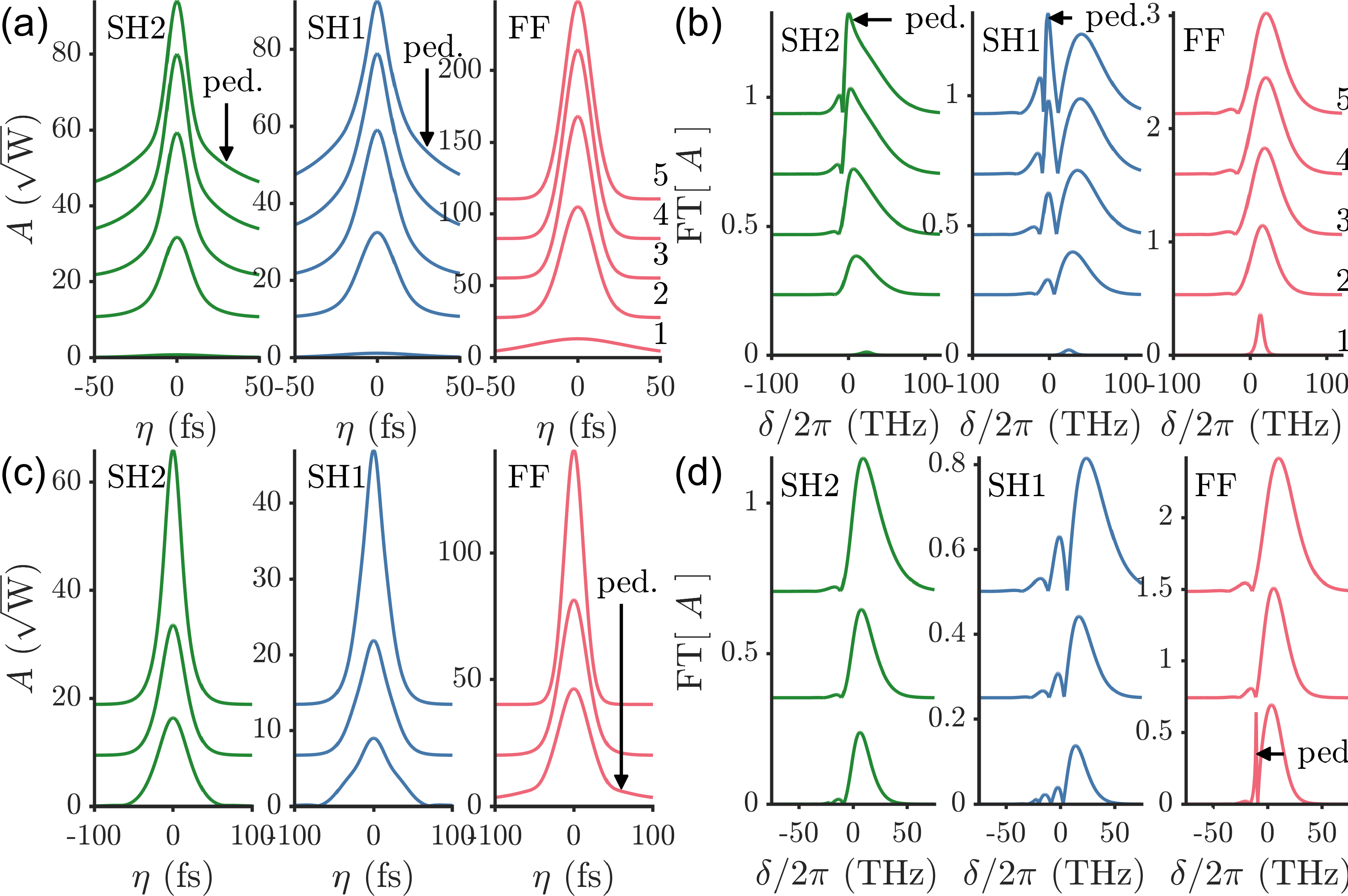}
    \caption{ Soliton profiles in the time ((a) and (c)) and frequency domain ((b) and (d)). (a) and (b) show solitons on the solid line as shown in figure \ref{c:mode_crossing:f:soliton_analysis} (a) labeled with the same numbers. (c) and (d) show solitons on the lower half of the dashed line in figure \ref{c:mode_crossing:f:soliton_analysis} (a).  FF, SH1 and SH2 are shown in panels from right to left in red, blue and green respectively. Evidence of pedestal formation is labeled with 'ped.'. }
    \label{c:mode_crossing:f:soliton_profiles}
\end{figure*}

To explain changes of the solitons structure across their domain of existence, it is instructive to consider solutions along two constant $\nu$ lines as shown in Fig.~\ref{c:mode_crossing:f:soliton_analysis}(a). These lines are chosen as examples that allow us to discuss three important trends we observe in the structure of these solitons more generally as a function of $\mu$. The first of these occurs in the $\nu=-\SI{115}{fs/mm}$ (solid) line as $\mu$ approaches its minimum value, $\mu_{min}$, where we observe a rapid drop in the total energy of each component as shown in Fig.~\ref{c:mode_crossing:f:soliton_analysis}(b). This  drop occurs as the soliton broadens and drops in peak power as shown in Fig.~\ref{c:mode_crossing:f:soliton_profiles}(a). In this limiting case as power drops to zero the soliton becomes a linear wave packet. This behaviour is consistent with that of a Kerr soliton as $\mu \xrightarrow{} \mu_{min}$, for this reason we will refer to this case a 'Kerr-like'.

The next case of interest is seen in the $\nu=-\SI{62}{fs/mm}$ (dashed) line as  $\mu \xrightarrow{} \mu_{min}$. In this case we see a small uptick in energy in the FF component shown in Fig.~\ref{c:mode_crossing:f:soliton_analysis}(b). The reason for this can be seen in the soliton profiles in Fig.~\ref{c:mode_crossing:f:soliton_profiles}(c) where we see a pedestal beginning to form in the FF component as it approaches $\mu_{min}$. We will call this case 'FF-pedestal'. We also point out that the soliton is not broadening and decreasing in power as in the 'Kerr-like' case. 

The final case occurs in both constant $\nu$ lines where we see an uptick in energy in the SH components as the solitons approach their maximal $\mu$ values, $\mu_{max}$. Once again, we see the formation of pedestals, this time in the SH components and we give this case the name 'SH-pedestal'.

The onset of these pedestals can also be observed in the frequency domain where sharp peaks begin to form as in Fig.~\ref{c:mode_crossing:f:soliton_profiles}(b) and (d). These peaks are detuned from the main core of the soliton, as expected from our previous observation of distinct core and tail frequencies in Fig.~\ref{c:mode_crossing:f:soliton_analysis}(c). Using both the core and tail frequency detunings for $\delta$, the two constant $\nu$ lines are mapped onto the $\mucw$-$\delta$ plane in Figs.~\ref{c:mode_crossing:f:soliton_analysis}(c) and (d), respectively. These data show very different behaviour of the core and tail frequencies as a function of $\mucw$.  The core frequencies change smoothly and remain in the region of CW solution existence. The tail frequencies change abruptly as different frequency components in the soliton tail become dominant. We also point out that these tail frequency data approach the linear dispersion lines at exactly the points predicted by our analysis in Eqs.~\eqref{c:mode_crossing:e:delta_tails}.

As $\mu \xrightarrow{} \mu_{min}$ we see the core and tail frequencies of the $\nu=-\SI{115}{fs/mm}$ (solid) line approach the same point on the FF linear dispersion. This is not true for the $\nu=-\SI{62}{fs/mm}$ (dashed) line or either line as $\mu \rightarrow \mu_{max}$. We believe the difference between these cases comes from the approach of the soliton to the boundary of CW solution existence. As the $\nu=-\SI{62}{fs/mm}$ (dashed) line approaches the FF linear dispersion in its tail frequency, the shift of the soliton body is frustrated by the boundary of CW solution existence. The soliton core frequency cannot cross this boundary as CW solution existence is a necessary criteria for soliton existence. The same behaviour occurs in both lines as $\mu \rightarrow \mu_{max}$.

\begin{figure*}
    \centering
    \includegraphics{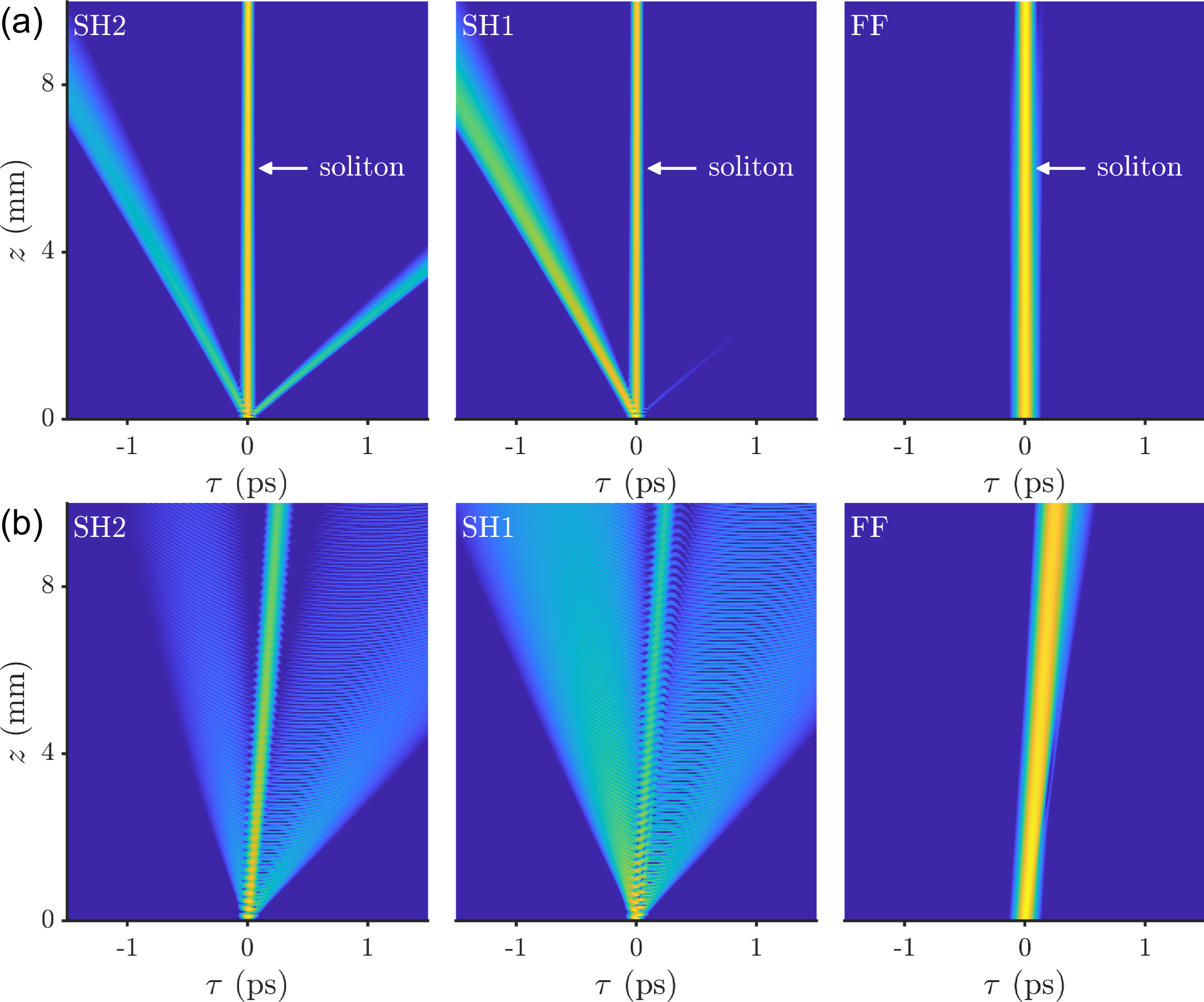}
    \caption{ Simulations of soliton generation from simple initial pulses in the FF component plotted in the time domain. Initial pulses had sech field shape with 33fs FWHM and 160W peak power.
     Frequency detuning (a) $\delta/2\pi = \SI{13}{THz}$ ($\lambda \approx \SI{1200}{nm}$) and  
    (b) $\delta/2\pi = \SI{2.9}{THz}$ ($\lambda \approx \SI{1230}{nm}$). }
    \label{ c:mode_crossing:f:sol_sims }
\end{figure*}

Finally,we used numerical methods to simulate pulse propagation in our system. Our first step was to simulate propagation of numerical soliton solutions. We found these to be generally stable over distances of tens of millimeters, which are feasible maximum lengths for such LN nanowaveguides. Next, we simulated the generation of solitons from simple initial pulses, similar to what might be feasible experimentally. In our simulations we found that solitons could be generated from a single sech shaped pulse in the FF component. One example of such a simulation is shown in Fig.~\ref{ c:mode_crossing:f:sol_sims }(a). In this example, formation of a soliton is clearly visible after $~1$mm of propagation, suggesting a minimum waveguide length for any such experimental investigation.

According to our results presented in Fig.~\ref{c:mode_crossing:f:cw}(a), solitons are predicted only for certain frequency detunings $\delta$. Our simulations showed soliton generation was less favourable when shifting the initial pulse away from this predicted region. An example simulation initialised with a FF pulse outside the soliton existence region is shown in Fig.~\ref{ c:mode_crossing:f:sol_sims }(b). Temporal broadening in all three spectral components is clearly visible, and the dynamics is very different from the soliton generated in panel (a). 

\section{Summary}

 We have analysed nonlinear CW and soliton solutions in a $\chitwo$ waveguide with two guided modes making an avoided crossing at the SH frequency. We have developed a model with two linearly coupled "pure" modes describing the dispersion in the avoided crossing region.
 Introducing a difference in effective nonlinearity between each of the two pure modes and the FF mode, we also reproduced the dispersion of nonlinearity in the system. We find that CW solutions
 generally exist within four domains in the plane of their two parameters: the propagation constant, $\mucw$, and the frequency detuning, $\delta$. Linear dispersion of the FF and the two SH modes, and the additional condition $\mucw=\musin(\delta)$ where CW solutions become singular, together define the boundaries of the domains.  
 Our analysis shows CW solutions to be unstable with respect to modulation over the vast majority of their existence domains. In some cases, simulated propagation of CW solutions with small initial perturbation revealed formation of trains of solitons, as the result of the instability development.
 We also derived existence conditions of the three-component soliton solutions in our system, and obtained corresponding numerical solutions. Our analysis reveals that solitons generally have frequency chirp in all three components, with the core of the soliton and its tails generally having distinct frequencies.
 Close to their existence domain boundaries, these solitons show signs of pedestal formation in either SH or FF components, when their tail frequencies reside in regions of no CW solution existence. This pedestal formation is accompanied by the formation of characteristic spectral features. Finally, we simulated soliton propagation where we found them to be stable over distances typical for the lengths of LN nano-waveguides.  Generation of solitons from simple, low-power pulses in the FF component was also successfully simulated, suggesting the potential for experimental studies of soliton dynamics in such systems.

\section{Acknowledgements}
We thank K. Gallo and H. Fergestad for enlightening discussions. WRR acknowledges funding and support from the U.K. Engineering and Physical Sciences Research Council (EPSRC) Centre for Doctoral Training in Condensed Matter Physics (CDTCMP), Grant No. EP/L015544/1.

\bibliography{mendeley.bib}

\begin{thebibliography}{28}%
\makeatletter
\providecommand \@ifxundefined [1]{%
 \@ifx{#1\undefined}
}%
\providecommand \@ifnum [1]{%
 \ifnum #1\expandafter \@firstoftwo
 \else \expandafter \@secondoftwo
 \fi
}%
\providecommand \@ifx [1]{%
 \ifx #1\expandafter \@firstoftwo
 \else \expandafter \@secondoftwo
 \fi
}%
\providecommand \natexlab [1]{#1}%
\providecommand \enquote  [1]{``#1''}%
\providecommand \bibnamefont  [1]{#1}%
\providecommand \bibfnamefont [1]{#1}%
\providecommand \citenamefont [1]{#1}%
\providecommand \href@noop [0]{\@secondoftwo}%
\providecommand \href [0]{\begingroup \@sanitize@url \@href}%
\providecommand \@href[1]{\@@startlink{#1}\@@href}%
\providecommand \@@href[1]{\endgroup#1\@@endlink}%
\providecommand \@sanitize@url [0]{\catcode `\\12\catcode `\$12\catcode
  `\&12\catcode `\#12\catcode `\^12\catcode `\_12\catcode `\%12\relax}%
\providecommand \@@startlink[1]{}%
\providecommand \@@endlink[0]{}%
\providecommand \url  [0]{\begingroup\@sanitize@url \@url }%
\providecommand \@url [1]{\endgroup\@href {#1}{\urlprefix }}%
\providecommand \urlprefix  [0]{URL }%
\providecommand \Eprint [0]{\href }%
\providecommand \doibase [0]{http://dx.doi.org/}%
\providecommand \selectlanguage [0]{\@gobble}%
\providecommand \bibinfo  [0]{\@secondoftwo}%
\providecommand \bibfield  [0]{\@secondoftwo}%
\providecommand \translation [1]{[#1]}%
\providecommand \BibitemOpen [0]{}%
\providecommand \bibitemStop [0]{}%
\providecommand \bibitemNoStop [0]{.\EOS\space}%
\providecommand \EOS [0]{\spacefactor3000\relax}%
\providecommand \BibitemShut  [1]{\csname bibitem#1\endcsname}%
\let\auto@bib@innerbib\@empty
\bibitem [{\citenamefont {Kowligy}\ \emph {et~al.}(2018)\citenamefont
  {Kowligy}, \citenamefont {Lind}, \citenamefont {Hickstein}, \citenamefont
  {Carlson}, \citenamefont {Timmers}, \citenamefont {Nader}, \citenamefont
  {Cruz}, \citenamefont {Ycas}, \citenamefont {Papp},\ and\ \citenamefont
  {Diddams}}]{Kowligy2018Mid-infraredWaveguides}%
  \BibitemOpen
  \bibfield  {author} {\bibinfo {author} {\bibfnamefont {A.~S.}\ \bibnamefont
  {Kowligy}}, \bibinfo {author} {\bibfnamefont {A.}~\bibnamefont {Lind}},
  \bibinfo {author} {\bibfnamefont {D.~D.}\ \bibnamefont {Hickstein}}, \bibinfo
  {author} {\bibfnamefont {D.~R.}\ \bibnamefont {Carlson}}, \bibinfo {author}
  {\bibfnamefont {H.}~\bibnamefont {Timmers}}, \bibinfo {author} {\bibfnamefont
  {N.}~\bibnamefont {Nader}}, \bibinfo {author} {\bibfnamefont {F.~C.}\
  \bibnamefont {Cruz}}, \bibinfo {author} {\bibfnamefont {G.}~\bibnamefont
  {Ycas}}, \bibinfo {author} {\bibfnamefont {S.~B.}\ \bibnamefont {Papp}}, \
  and\ \bibinfo {author} {\bibfnamefont {S.~A.}\ \bibnamefont {Diddams}},\
  }\href {\doibase 10.1364/OL.43.001678} {\bibfield  {journal} {\bibinfo
  {journal} {Optics Letters}\ }\textbf {\bibinfo {volume} {43}},\ \bibinfo
  {pages} {1678} (\bibinfo {year} {2018})}\BibitemShut {NoStop}%
\bibitem [{\citenamefont {Guo}\ \emph {et~al.}(2015)\citenamefont {Guo},
  \citenamefont {Zhou}, \citenamefont {Steinert}, \citenamefont {Setzpfandt},
  \citenamefont {Pertsch}, \citenamefont {Chung}, \citenamefont {Chen},\ and\
  \citenamefont {Bache}}]{Guo2015SupercontinuumMatching}%
  \BibitemOpen
  \bibfield  {author} {\bibinfo {author} {\bibfnamefont {H.}~\bibnamefont
  {Guo}}, \bibinfo {author} {\bibfnamefont {B.}~\bibnamefont {Zhou}}, \bibinfo
  {author} {\bibfnamefont {M.}~\bibnamefont {Steinert}}, \bibinfo {author}
  {\bibfnamefont {F.}~\bibnamefont {Setzpfandt}}, \bibinfo {author}
  {\bibfnamefont {T.}~\bibnamefont {Pertsch}}, \bibinfo {author} {\bibfnamefont
  {H.-p.}\ \bibnamefont {Chung}}, \bibinfo {author} {\bibfnamefont {Y.-H.}\
  \bibnamefont {Chen}}, \ and\ \bibinfo {author} {\bibfnamefont
  {M.}~\bibnamefont {Bache}},\ }\href {\doibase 10.1364/OL.40.000629}
  {\bibfield  {journal} {\bibinfo  {journal} {Optics Letters}\ }\textbf
  {\bibinfo {volume} {40}},\ \bibinfo {pages} {629} (\bibinfo {year}
  {2015})}\BibitemShut {NoStop}%
\bibitem [{\citenamefont {Rowe}\ \emph {et~al.}(2019)\citenamefont {Rowe},
  \citenamefont {Skryabin},\ and\ \citenamefont
  {Gorbach}}]{Rowe2019TemporalNanowaveguides}%
  \BibitemOpen
  \bibfield  {author} {\bibinfo {author} {\bibfnamefont {W.~R.}\ \bibnamefont
  {Rowe}}, \bibinfo {author} {\bibfnamefont {D.~V.}\ \bibnamefont {Skryabin}},
  \ and\ \bibinfo {author} {\bibfnamefont {A.~V.}\ \bibnamefont {Gorbach}},\
  }\href {\doibase 10.1103/PhysRevResearch.1.033146} {\bibfield  {journal}
  {\bibinfo  {journal} {Physical Review Research}\ }\textbf {\bibinfo {volume}
  {1}},\ \bibinfo {pages} {033146} (\bibinfo {year} {2019})}\BibitemShut
  {NoStop}%
\bibitem [{\citenamefont {Rowe}\ \emph {et~al.}(2020)\citenamefont {Rowe},
  \citenamefont {Skryabin},\ and\ \citenamefont
  {Gorbach}}]{Rowe2020RamanNonlinearities}%
  \BibitemOpen
  \bibfield  {author} {\bibinfo {author} {\bibfnamefont {W.~R.}\ \bibnamefont
  {Rowe}}, \bibinfo {author} {\bibfnamefont {D.~V.}\ \bibnamefont {Skryabin}},
  \ and\ \bibinfo {author} {\bibfnamefont {A.~V.}\ \bibnamefont {Gorbach}},\
  }\href {\doibase 10.1103/PhysRevA.102.023523} {\bibfield  {journal} {\bibinfo
   {journal} {Physical Review A}\ }\textbf {\bibinfo {volume} {102}},\ \bibinfo
  {pages} {23523} (\bibinfo {year} {2020})}\BibitemShut {NoStop}%
\bibitem [{\citenamefont {Poberaj}\ \emph {et~al.}(2012)\citenamefont
  {Poberaj}, \citenamefont {Hu}, \citenamefont {Sohler},\ and\ \citenamefont
  {G{\"{u}}nter}}]{Poberaj2012LithiumDevices}%
  \BibitemOpen
  \bibfield  {author} {\bibinfo {author} {\bibfnamefont {G.}~\bibnamefont
  {Poberaj}}, \bibinfo {author} {\bibfnamefont {H.}~\bibnamefont {Hu}},
  \bibinfo {author} {\bibfnamefont {W.}~\bibnamefont {Sohler}}, \ and\ \bibinfo
  {author} {\bibfnamefont {P.}~\bibnamefont {G{\"{u}}nter}},\ }\href {\doibase
  10.1002/lpor.201100035} {\bibfield  {journal} {\bibinfo  {journal} {Laser
  {\&} Photonics Reviews}\ }\textbf {\bibinfo {volume} {6}},\ \bibinfo {pages}
  {488} (\bibinfo {year} {2012})}\BibitemShut {NoStop}%
\bibitem [{\citenamefont {Boes}\ \emph {et~al.}(2018)\citenamefont {Boes},
  \citenamefont {Corcoran}, \citenamefont {Chang}, \citenamefont {Bowers},\
  and\ \citenamefont {Mitchell}}]{Boes2018StatusCircuits}%
  \BibitemOpen
  \bibfield  {author} {\bibinfo {author} {\bibfnamefont {A.}~\bibnamefont
  {Boes}}, \bibinfo {author} {\bibfnamefont {B.}~\bibnamefont {Corcoran}},
  \bibinfo {author} {\bibfnamefont {L.}~\bibnamefont {Chang}}, \bibinfo
  {author} {\bibfnamefont {J.}~\bibnamefont {Bowers}}, \ and\ \bibinfo {author}
  {\bibfnamefont {A.}~\bibnamefont {Mitchell}},\ }\href {\doibase
  10.1002/lpor.201700256} {\bibfield  {journal} {\bibinfo  {journal} {Laser
  {\&} Photonics Reviews}\ }\textbf {\bibinfo {volume} {12}},\ \bibinfo {pages}
  {1700256} (\bibinfo {year} {2018})}\BibitemShut {NoStop}%
\bibitem [{\citenamefont {Huang}\ \emph {et~al.}(2015)\citenamefont {Huang},
  \citenamefont {Zhao}, \citenamefont {Kamyab}, \citenamefont {Rostami},
  \citenamefont {Capolino},\ and\ \citenamefont
  {Boyraz}}]{Huang2015Sub-micronLithography}%
  \BibitemOpen
  \bibfield  {author} {\bibinfo {author} {\bibfnamefont {Y.}~\bibnamefont
  {Huang}}, \bibinfo {author} {\bibfnamefont {Q.}~\bibnamefont {Zhao}},
  \bibinfo {author} {\bibfnamefont {L.}~\bibnamefont {Kamyab}}, \bibinfo
  {author} {\bibfnamefont {A.}~\bibnamefont {Rostami}}, \bibinfo {author}
  {\bibfnamefont {F.}~\bibnamefont {Capolino}}, \ and\ \bibinfo {author}
  {\bibfnamefont {O.}~\bibnamefont {Boyraz}},\ }\href {\doibase
  10.1364/OE.23.006780} {\bibfield  {journal} {\bibinfo  {journal} {Optics
  Express}\ }\textbf {\bibinfo {volume} {23}},\ \bibinfo {pages} {6780}
  (\bibinfo {year} {2015})}\BibitemShut {NoStop}%
\bibitem [{\citenamefont {Wilson}\ \emph {et~al.}(2020)\citenamefont {Wilson},
  \citenamefont {Schneider}, \citenamefont {H{\"{o}}nl}, \citenamefont
  {Anderson}, \citenamefont {Baumgartner}, \citenamefont {Czornomaz},
  \citenamefont {Kippenberg},\ and\ \citenamefont
  {Seidler}}]{Wilson2020IntegratedPhotonics}%
  \BibitemOpen
  \bibfield  {author} {\bibinfo {author} {\bibfnamefont {D.~J.}\ \bibnamefont
  {Wilson}}, \bibinfo {author} {\bibfnamefont {K.}~\bibnamefont {Schneider}},
  \bibinfo {author} {\bibfnamefont {S.}~\bibnamefont {H{\"{o}}nl}}, \bibinfo
  {author} {\bibfnamefont {M.}~\bibnamefont {Anderson}}, \bibinfo {author}
  {\bibfnamefont {Y.}~\bibnamefont {Baumgartner}}, \bibinfo {author}
  {\bibfnamefont {L.}~\bibnamefont {Czornomaz}}, \bibinfo {author}
  {\bibfnamefont {T.~J.}\ \bibnamefont {Kippenberg}}, \ and\ \bibinfo {author}
  {\bibfnamefont {P.}~\bibnamefont {Seidler}},\ }\href {\doibase
  10.1038/s41566-019-0537-9} {\bibfield  {journal} {\bibinfo  {journal} {Nature
  Photonics}\ }\textbf {\bibinfo {volume} {14}},\ \bibinfo {pages} {57}
  (\bibinfo {year} {2020})}\BibitemShut {NoStop}%
\bibitem [{\citenamefont {Zeng}\ \emph {et~al.}(2008)\citenamefont {Zeng},
  \citenamefont {Ashihara}, \citenamefont {Chen}, \citenamefont {Shimura},\
  and\ \citenamefont {Kuroda}}]{Zeng2008Two-colorNiobate}%
  \BibitemOpen
  \bibfield  {author} {\bibinfo {author} {\bibfnamefont {X.}~\bibnamefont
  {Zeng}}, \bibinfo {author} {\bibfnamefont {S.}~\bibnamefont {Ashihara}},
  \bibinfo {author} {\bibfnamefont {X.}~\bibnamefont {Chen}}, \bibinfo {author}
  {\bibfnamefont {T.}~\bibnamefont {Shimura}}, \ and\ \bibinfo {author}
  {\bibfnamefont {K.}~\bibnamefont {Kuroda}},\ }\href {\doibase
  10.1016/j.optcom.2008.04.080} {\bibfield  {journal} {\bibinfo  {journal}
  {Optics Communications}\ }\textbf {\bibinfo {volume} {281}},\ \bibinfo
  {pages} {4499} (\bibinfo {year} {2008})}\BibitemShut {NoStop}%
\bibitem [{\citenamefont {Bache}\ and\ \citenamefont
  {Wise}(2010)}]{Bache2010Type-ILasers}%
  \BibitemOpen
  \bibfield  {author} {\bibinfo {author} {\bibfnamefont {M.}~\bibnamefont
  {Bache}}\ and\ \bibinfo {author} {\bibfnamefont {F.~W.}\ \bibnamefont
  {Wise}},\ }\href {\doibase 10.1103/PhysRevA.81.053815} {\bibfield  {journal}
  {\bibinfo  {journal} {Physical Review A}\ }\textbf {\bibinfo {volume} {81}},\
  \bibinfo {pages} {053815} (\bibinfo {year} {2010})}\BibitemShut {NoStop}%
\bibitem [{\citenamefont {Zeng}\ \emph {et~al.}(2012)\citenamefont {Zeng},
  \citenamefont {Guo}, \citenamefont {Zhou},\ and\ \citenamefont
  {Bache}}]{Zeng2012SolitonNonlinearities}%
  \BibitemOpen
  \bibfield  {author} {\bibinfo {author} {\bibfnamefont {X.}~\bibnamefont
  {Zeng}}, \bibinfo {author} {\bibfnamefont {H.}~\bibnamefont {Guo}}, \bibinfo
  {author} {\bibfnamefont {B.}~\bibnamefont {Zhou}}, \ and\ \bibinfo {author}
  {\bibfnamefont {M.}~\bibnamefont {Bache}},\ }\href {\doibase
  10.1364/OE.20.027071} {\bibfield  {journal} {\bibinfo  {journal} {Optics
  Express}\ }\textbf {\bibinfo {volume} {20}},\ \bibinfo {pages} {27071}
  (\bibinfo {year} {2012})}\BibitemShut {NoStop}%
\bibitem [{\citenamefont {Yu}\ \emph {et~al.}(2020)\citenamefont {Yu},
  \citenamefont {Shao}, \citenamefont {Okawachi}, \citenamefont {Gaeta},\ and\
  \citenamefont {Loncar}}]{Yu2020UltravioletWaveguides}%
  \BibitemOpen
  \bibfield  {author} {\bibinfo {author} {\bibfnamefont {M.}~\bibnamefont
  {Yu}}, \bibinfo {author} {\bibfnamefont {L.}~\bibnamefont {Shao}}, \bibinfo
  {author} {\bibfnamefont {Y.}~\bibnamefont {Okawachi}}, \bibinfo {author}
  {\bibfnamefont {A.~L.}\ \bibnamefont {Gaeta}}, \ and\ \bibinfo {author}
  {\bibfnamefont {M.}~\bibnamefont {Loncar}},\ }in\ \href {\doibase
  10.1364/CLEO{\_}SI.2020.STu4H.1} {\emph {\bibinfo {booktitle} {Conference on
  Lasers and Electro-Optics}}},\ \bibinfo {series and number} {OSA Technical
  Digest}\ (\bibinfo  {publisher} {Optical Society of America},\ \bibinfo
  {address} {Washington, DC},\ \bibinfo {year} {2020})\ p.\ \bibinfo {pages}
  {STu4H.1}\BibitemShut {NoStop}%
\bibitem [{\citenamefont {Wang}\ \emph {et~al.}(2018)\citenamefont {Wang},
  \citenamefont {Langrock}, \citenamefont {Marandi}, \citenamefont {Jankowski},
  \citenamefont {Zhang}, \citenamefont {Desiatov}, \citenamefont {Fejer},\ and\
  \citenamefont {Lon{\v{c}}ar}}]{Wang2018Ultrahigh-efficiencyWaveguides}%
  \BibitemOpen
  \bibfield  {author} {\bibinfo {author} {\bibfnamefont {C.}~\bibnamefont
  {Wang}}, \bibinfo {author} {\bibfnamefont {C.}~\bibnamefont {Langrock}},
  \bibinfo {author} {\bibfnamefont {A.}~\bibnamefont {Marandi}}, \bibinfo
  {author} {\bibfnamefont {M.}~\bibnamefont {Jankowski}}, \bibinfo {author}
  {\bibfnamefont {M.}~\bibnamefont {Zhang}}, \bibinfo {author} {\bibfnamefont
  {B.}~\bibnamefont {Desiatov}}, \bibinfo {author} {\bibfnamefont {M.~M.}\
  \bibnamefont {Fejer}}, \ and\ \bibinfo {author} {\bibfnamefont
  {M.}~\bibnamefont {Lon{\v{c}}ar}},\ }\href {\doibase 10.1364/OPTICA.5.001438}
  {\bibfield  {journal} {\bibinfo  {journal} {Optica}\ }\textbf {\bibinfo
  {volume} {5}},\ \bibinfo {pages} {1438} (\bibinfo {year} {2018})}\BibitemShut
  {NoStop}%
\bibitem [{\citenamefont {Lu}\ \emph {et~al.}(2019)\citenamefont {Lu},
  \citenamefont {Moille}, \citenamefont {Li}, \citenamefont {Westly},
  \citenamefont {Singh}, \citenamefont {Rao}, \citenamefont {Yu}, \citenamefont
  {Briles}, \citenamefont {Papp},\ and\ \citenamefont
  {Srinivasan}}]{Lu2019EfficientNanophotonics}%
  \BibitemOpen
  \bibfield  {author} {\bibinfo {author} {\bibfnamefont {X.}~\bibnamefont
  {Lu}}, \bibinfo {author} {\bibfnamefont {G.}~\bibnamefont {Moille}}, \bibinfo
  {author} {\bibfnamefont {Q.}~\bibnamefont {Li}}, \bibinfo {author}
  {\bibfnamefont {D.~A.}\ \bibnamefont {Westly}}, \bibinfo {author}
  {\bibfnamefont {A.}~\bibnamefont {Singh}}, \bibinfo {author} {\bibfnamefont
  {A.}~\bibnamefont {Rao}}, \bibinfo {author} {\bibfnamefont {S.-P.}\
  \bibnamefont {Yu}}, \bibinfo {author} {\bibfnamefont {T.~C.}\ \bibnamefont
  {Briles}}, \bibinfo {author} {\bibfnamefont {S.~B.}\ \bibnamefont {Papp}}, \
  and\ \bibinfo {author} {\bibfnamefont {K.}~\bibnamefont {Srinivasan}},\
  }\href {\doibase 10.1038/s41566-019-0464-9} {\bibfield  {journal} {\bibinfo
  {journal} {Nature Photonics}\ }\textbf {\bibinfo {volume} {13}},\ \bibinfo
  {pages} {593} (\bibinfo {year} {2019})}\BibitemShut {NoStop}%
\bibitem [{\citenamefont {Zhao}\ \emph {et~al.}(2020)\citenamefont {Zhao},
  \citenamefont {R{\"{u}}sing}, \citenamefont {Javid}, \citenamefont {Ling},
  \citenamefont {Li}, \citenamefont {Lin},\ and\ \citenamefont
  {Mookherjea}}]{Zhao2020Shallow-etchedGeneration}%
  \BibitemOpen
  \bibfield  {author} {\bibinfo {author} {\bibfnamefont {J.}~\bibnamefont
  {Zhao}}, \bibinfo {author} {\bibfnamefont {M.}~\bibnamefont {R{\"{u}}sing}},
  \bibinfo {author} {\bibfnamefont {U.~A.}\ \bibnamefont {Javid}}, \bibinfo
  {author} {\bibfnamefont {J.}~\bibnamefont {Ling}}, \bibinfo {author}
  {\bibfnamefont {M.}~\bibnamefont {Li}}, \bibinfo {author} {\bibfnamefont
  {Q.}~\bibnamefont {Lin}}, \ and\ \bibinfo {author} {\bibfnamefont
  {S.}~\bibnamefont {Mookherjea}},\ }\href {\doibase 10.1364/OE.395545}
  {\bibfield  {journal} {\bibinfo  {journal} {Optics Express}\ }\textbf
  {\bibinfo {volume} {28}},\ \bibinfo {pages} {19669} (\bibinfo {year}
  {2020})}\BibitemShut {NoStop}%
\bibitem [{\citenamefont {Skryabin}\ and\ \citenamefont
  {Gorbach}(2010)}]{Skryabin2010TheoryWaves}%
  \BibitemOpen
  \bibfield  {author} {\bibinfo {author} {\bibfnamefont {D.~V.}\ \bibnamefont
  {Skryabin}}\ and\ \bibinfo {author} {\bibfnamefont {A.~V.}\ \bibnamefont
  {Gorbach}},\ }in\ \href {\doibase DOI: 10.1017/CBO9780511750465.010} {\emph
  {\bibinfo {booktitle} {Supercontinuum Generation in Optical Fibers}}},\
  \bibinfo {editor} {edited by\ \bibinfo {editor} {\bibfnamefont {J.~M.}\
  \bibnamefont {Dudley}}\ and\ \bibinfo {editor} {\bibfnamefont {J.~R.}\
  \bibnamefont {Taylor}}}\ (\bibinfo  {publisher} {Cambridge University
  Press},\ \bibinfo {address} {Cambridge},\ \bibinfo {year} {2010})\ pp.\
  \bibinfo {pages} {178--198}\BibitemShut {NoStop}%
\bibitem [{\citenamefont {Zelmon}\ \emph {et~al.}(1997)\citenamefont {Zelmon},
  \citenamefont {Small},\ and\ \citenamefont
  {Jundt}}]{Zelmon1997InfraredNiobate}%
  \BibitemOpen
  \bibfield  {author} {\bibinfo {author} {\bibfnamefont {D.~E.}\ \bibnamefont
  {Zelmon}}, \bibinfo {author} {\bibfnamefont {D.~L.}\ \bibnamefont {Small}}, \
  and\ \bibinfo {author} {\bibfnamefont {D.}~\bibnamefont {Jundt}},\ }\href
  {\doibase 10.1364/JOSAB.14.003319} {\bibfield  {journal} {\bibinfo  {journal}
  {Journal of the Optical Society of America B}\ }\textbf {\bibinfo {volume}
  {14}},\ \bibinfo {pages} {3319} (\bibinfo {year} {1997})}\BibitemShut
  {NoStop}%
\bibitem [{\citenamefont {Zhu}\ \emph {et~al.}(2021)\citenamefont {Zhu},
  \citenamefont {Shao}, \citenamefont {Yu}, \citenamefont {Cheng},
  \citenamefont {Desiatov}, \citenamefont {Xin}, \citenamefont {Hu},
  \citenamefont {Holzgrafe}, \citenamefont {Ghosh}, \citenamefont
  {Shams-Ansari}, \citenamefont {Puma}, \citenamefont {Sinclair}, \citenamefont
  {Reimer}, \citenamefont {Zhang},\ and\ \citenamefont
  {Lon{\v{c}}ar}}]{Zhu2021IntegratedNiobate}%
  \BibitemOpen
  \bibfield  {author} {\bibinfo {author} {\bibfnamefont {D.}~\bibnamefont
  {Zhu}}, \bibinfo {author} {\bibfnamefont {L.}~\bibnamefont {Shao}}, \bibinfo
  {author} {\bibfnamefont {M.}~\bibnamefont {Yu}}, \bibinfo {author}
  {\bibfnamefont {R.}~\bibnamefont {Cheng}}, \bibinfo {author} {\bibfnamefont
  {B.}~\bibnamefont {Desiatov}}, \bibinfo {author} {\bibfnamefont {C.~J.}\
  \bibnamefont {Xin}}, \bibinfo {author} {\bibfnamefont {Y.}~\bibnamefont
  {Hu}}, \bibinfo {author} {\bibfnamefont {J.}~\bibnamefont {Holzgrafe}},
  \bibinfo {author} {\bibfnamefont {S.}~\bibnamefont {Ghosh}}, \bibinfo
  {author} {\bibfnamefont {A.}~\bibnamefont {Shams-Ansari}}, \bibinfo {author}
  {\bibfnamefont {E.}~\bibnamefont {Puma}}, \bibinfo {author} {\bibfnamefont
  {N.}~\bibnamefont {Sinclair}}, \bibinfo {author} {\bibfnamefont
  {C.}~\bibnamefont {Reimer}}, \bibinfo {author} {\bibfnamefont
  {M.}~\bibnamefont {Zhang}}, \ and\ \bibinfo {author} {\bibfnamefont
  {M.}~\bibnamefont {Lon{\v{c}}ar}},\ }\href {\doibase 10.1364/AOP.411024}
  {\bibfield  {journal} {\bibinfo  {journal} {Advances in Optics and
  Photonics}\ }\textbf {\bibinfo {volume} {13}},\ \bibinfo {pages} {242}
  (\bibinfo {year} {2021})}\BibitemShut {NoStop}%
\bibitem [{\citenamefont {Cai}\ \emph {et~al.}(2018)\citenamefont {Cai},
  \citenamefont {Gorbach}, \citenamefont {Wang}, \citenamefont {Hu},\ and\
  \citenamefont {Ding}}]{Cai2018HighlyNano-waveguides}%
  \BibitemOpen
  \bibfield  {author} {\bibinfo {author} {\bibfnamefont {L.}~\bibnamefont
  {Cai}}, \bibinfo {author} {\bibfnamefont {A.~V.}\ \bibnamefont {Gorbach}},
  \bibinfo {author} {\bibfnamefont {Y.}~\bibnamefont {Wang}}, \bibinfo {author}
  {\bibfnamefont {H.}~\bibnamefont {Hu}}, \ and\ \bibinfo {author}
  {\bibfnamefont {W.}~\bibnamefont {Ding}},\ }\href {\doibase
  10.1038/s41598-018-31017-0} {\bibfield  {journal} {\bibinfo  {journal}
  {Scientific Reports}\ }\textbf {\bibinfo {volume} {8}},\ \bibinfo {pages}
  {12478} (\bibinfo {year} {2018})}\BibitemShut {NoStop}%
\bibitem [{\citenamefont {Skryabin}(2004)}]{Skryabin2004CoupledFibers}%
  \BibitemOpen
  \bibfield  {author} {\bibinfo {author} {\bibfnamefont {D.~V.}\ \bibnamefont
  {Skryabin}},\ }\href {\doibase 10.1364/opex.12.004841} {\bibfield  {journal}
  {\bibinfo  {journal} {Optics Express}\ }\textbf {\bibinfo {volume} {12}},\
  \bibinfo {pages} {4841} (\bibinfo {year} {2004})}\BibitemShut {NoStop}%
\bibitem [{\citenamefont {Tani}\ \emph {et~al.}(2018)\citenamefont {Tani},
  \citenamefont {K{\"{o}}ttig}, \citenamefont {Novoa}, \citenamefont {Keding},\
  and\ \citenamefont {Russell}}]{Tani2018EffectFibers}%
  \BibitemOpen
  \bibfield  {author} {\bibinfo {author} {\bibfnamefont {F.}~\bibnamefont
  {Tani}}, \bibinfo {author} {\bibfnamefont {F.}~\bibnamefont {K{\"{o}}ttig}},
  \bibinfo {author} {\bibfnamefont {D.}~\bibnamefont {Novoa}}, \bibinfo
  {author} {\bibfnamefont {R.}~\bibnamefont {Keding}}, \ and\ \bibinfo {author}
  {\bibfnamefont {P.~S.}\ \bibnamefont {Russell}},\ }\href {\doibase
  10.1364/PRJ.6.000084} {\bibfield  {journal} {\bibinfo  {journal} {Photonics
  Research}\ }\textbf {\bibinfo {volume} {6}},\ \bibinfo {pages} {84} (\bibinfo
  {year} {2018})}\BibitemShut {NoStop}%
\bibitem [{\citenamefont {Yulin}\ \emph {et~al.}(2005)\citenamefont {Yulin},
  \citenamefont {Skryabin},\ and\ \citenamefont
  {Russell}}]{Yulin2005DissipativeFilms}%
  \BibitemOpen
  \bibfield  {author} {\bibinfo {author} {\bibfnamefont {A.~V.}\ \bibnamefont
  {Yulin}}, \bibinfo {author} {\bibfnamefont {D.~V.}\ \bibnamefont {Skryabin}},
  \ and\ \bibinfo {author} {\bibfnamefont {P.}~\bibnamefont {Russell}},\ }\href
  {\doibase 10.1364/OPEX.13.003529} {\bibfield  {journal} {\bibinfo  {journal}
  {Optics Express}\ }\textbf {\bibinfo {volume} {13}},\ \bibinfo {pages} {3529}
  (\bibinfo {year} {2005})}\BibitemShut {NoStop}%
\bibitem [{\citenamefont {Liu}\ \emph {et~al.}(2014)\citenamefont {Liu},
  \citenamefont {Xuan}, \citenamefont {Xue}, \citenamefont {Wang},
  \citenamefont {Chen}, \citenamefont {Metcalf}, \citenamefont {Wang},
  \citenamefont {Leaird}, \citenamefont {Qi},\ and\ \citenamefont
  {Weiner}}]{Liu2014InvestigationGeneration}%
  \BibitemOpen
  \bibfield  {author} {\bibinfo {author} {\bibfnamefont {Y.}~\bibnamefont
  {Liu}}, \bibinfo {author} {\bibfnamefont {Y.}~\bibnamefont {Xuan}}, \bibinfo
  {author} {\bibfnamefont {X.}~\bibnamefont {Xue}}, \bibinfo {author}
  {\bibfnamefont {P.-H.}\ \bibnamefont {Wang}}, \bibinfo {author}
  {\bibfnamefont {S.}~\bibnamefont {Chen}}, \bibinfo {author} {\bibfnamefont
  {A.~J.}\ \bibnamefont {Metcalf}}, \bibinfo {author} {\bibfnamefont
  {J.}~\bibnamefont {Wang}}, \bibinfo {author} {\bibfnamefont {D.~E.}\
  \bibnamefont {Leaird}}, \bibinfo {author} {\bibfnamefont {M.}~\bibnamefont
  {Qi}}, \ and\ \bibinfo {author} {\bibfnamefont {A.~M.}\ \bibnamefont
  {Weiner}},\ }\href {\doibase 10.1364/OPTICA.1.000137} {\bibfield  {journal}
  {\bibinfo  {journal} {Optica}\ }\textbf {\bibinfo {volume} {1}},\ \bibinfo
  {pages} {137} (\bibinfo {year} {2014})}\BibitemShut {NoStop}%
\bibitem [{\citenamefont {Herr}\ \emph {et~al.}(2014)\citenamefont {Herr},
  \citenamefont {Brasch}, \citenamefont {Jost}, \citenamefont {Mirgorodskiy},
  \citenamefont {Lihachev}, \citenamefont {Gorodetsky},\ and\ \citenamefont
  {Kippenberg}}]{Herr2014ModeMicroresonators}%
  \BibitemOpen
  \bibfield  {author} {\bibinfo {author} {\bibfnamefont {T.}~\bibnamefont
  {Herr}}, \bibinfo {author} {\bibfnamefont {V.}~\bibnamefont {Brasch}},
  \bibinfo {author} {\bibfnamefont {J.}~\bibnamefont {Jost}}, \bibinfo {author}
  {\bibfnamefont {I.}~\bibnamefont {Mirgorodskiy}}, \bibinfo {author}
  {\bibfnamefont {G.}~\bibnamefont {Lihachev}}, \bibinfo {author}
  {\bibfnamefont {M.}~\bibnamefont {Gorodetsky}}, \ and\ \bibinfo {author}
  {\bibfnamefont {T.}~\bibnamefont {Kippenberg}},\ }\href {\doibase
  10.1103/PhysRevLett.113.123901} {\bibfield  {journal} {\bibinfo  {journal}
  {Physical Review Letters}\ }\textbf {\bibinfo {volume} {113}},\ \bibinfo
  {pages} {123901} (\bibinfo {year} {2014})}\BibitemShut {NoStop}%
\bibitem [{\citenamefont {Wang}\ \emph {et~al.}(2020)\citenamefont {Wang},
  \citenamefont {Lu}, \citenamefont {Wu}, \citenamefont {Oh}, \citenamefont
  {Shen}, \citenamefont {Lee},\ and\ \citenamefont
  {Vahala}}]{Wang2020DiracMicroresonators}%
  \BibitemOpen
  \bibfield  {author} {\bibinfo {author} {\bibfnamefont {H.}~\bibnamefont
  {Wang}}, \bibinfo {author} {\bibfnamefont {Y.-K.}\ \bibnamefont {Lu}},
  \bibinfo {author} {\bibfnamefont {L.}~\bibnamefont {Wu}}, \bibinfo {author}
  {\bibfnamefont {D.~Y.}\ \bibnamefont {Oh}}, \bibinfo {author} {\bibfnamefont
  {B.}~\bibnamefont {Shen}}, \bibinfo {author} {\bibfnamefont {S.~H.}\
  \bibnamefont {Lee}}, \ and\ \bibinfo {author} {\bibfnamefont
  {K.}~\bibnamefont {Vahala}},\ }\href {\doibase 10.1038/s41377-020-00438-w}
  {\bibfield  {journal} {\bibinfo  {journal} {Light: Science {\&}
  Applications}\ }\textbf {\bibinfo {volume} {9}},\ \bibinfo {pages} {205}
  (\bibinfo {year} {2020})}\BibitemShut {NoStop}%
\bibitem [{\citenamefont {Conti}\ \emph {et~al.}(1998)\citenamefont {Conti},
  \citenamefont {Trillo},\ and\ \citenamefont
  {Assanto}}]{Conti1998TrappingSolitons}%
  \BibitemOpen
  \bibfield  {author} {\bibinfo {author} {\bibfnamefont {C.}~\bibnamefont
  {Conti}}, \bibinfo {author} {\bibfnamefont {S.}~\bibnamefont {Trillo}}, \
  and\ \bibinfo {author} {\bibfnamefont {G.}~\bibnamefont {Assanto}},\ }\href
  {\doibase 10.1364/ol.23.000334} {\bibfield  {journal} {\bibinfo  {journal}
  {Optics Letters}\ }\textbf {\bibinfo {volume} {23}},\ \bibinfo {pages} {334}
  (\bibinfo {year} {1998})}\BibitemShut {NoStop}%
\bibitem [{\citenamefont {Arraf}\ \emph {et~al.}(2001)\citenamefont {Arraf},
  \citenamefont {De~Sterke},\ and\ \citenamefont
  {He}}]{Arraf2001BrightNonlinearity}%
  \BibitemOpen
  \bibfield  {author} {\bibinfo {author} {\bibfnamefont {A.}~\bibnamefont
  {Arraf}}, \bibinfo {author} {\bibfnamefont {C.~M.}\ \bibnamefont
  {De~Sterke}}, \ and\ \bibinfo {author} {\bibfnamefont {H.}~\bibnamefont
  {He}},\ }\href {\doibase 10.1103/PhysRevE.63.026611} {\bibfield  {journal}
  {\bibinfo  {journal} {Physical Review E - Statistical, Nonlinear, and Soft
  Matter Physics}\ }\textbf {\bibinfo {volume} {63}},\ \bibinfo {pages} {1}
  (\bibinfo {year} {2001})}\BibitemShut {NoStop}%
\bibitem [{\citenamefont {Leitner}\ and\ \citenamefont
  {Malomed}(2005)}]{Leitner2005StabilityNonlinearity}%
  \BibitemOpen
  \bibfield  {author} {\bibinfo {author} {\bibfnamefont {Y.}~\bibnamefont
  {Leitner}}\ and\ \bibinfo {author} {\bibfnamefont {B.~A.}\ \bibnamefont
  {Malomed}},\ }\href {\doibase 10.1103/PhysRevE.71.057601} {\bibfield
  {journal} {\bibinfo  {journal} {Physical Review E - Statistical, Nonlinear,
  and Soft Matter Physics}\ }\textbf {\bibinfo {volume} {71}},\ \bibinfo
  {pages} {057601} (\bibinfo {year} {2005})}\BibitemShut {NoStop}%
\end{thebibliography}%

\end{document}